\documentclass[final,5p,times,twocolumn,sort&compress]{elsarticle}

\usepackage{mathtools}
\usepackage[usenames, dvipsnames]{color}

\newcommand*\circled[1]{{\small {\raisebox{.5pt}{\textcircled{\scriptsize{\raisebox{-.25pt} {#1}}}}}}}
\begin{document}
\frontmatter
\journal{Journal of Magnetic Resonance}
\title{Feedback control optimisation of ESR experiments\tnoteref{label1}}
\tnotetext[label1]{{\textcopyright} 2018. This manuscript version is available under the \textsc{cc-by-nc-nd} 4.0 license http://creativecommons.org/licenses/by-nc-nd/4.0/}

\author[soton,kit]{David L. Goodwin}
\author[caesr]{William K. Myers\corref{cor1}}
\ead{william.myers@chem.ox.ac.uk}
\author[caesr]{Christiane R. Timmel}
\author[soton]{Ilya Kuprov\corref{cor1}}
\ead{i.kuprov@soton.ac.uk}

\cortext[cor1]{Corresponding author}
\address[soton]{School of Chemistry, University of Southampton, Highfield Campus, Southampton SO17 1BJ, UK}
\address[kit]{Institute for Biological Interfaces 4 -- Magnetic Resonance, Karlsruhe Institute of Technology (KIT), Fritz-Haber-Weg 6, 76131 Karlsruhe, Germany}
\address[caesr]{Department of Chemistry, Centre for Advanced Electron Spin Resonance, University of Oxford, South Parks Road, Oxford OX1 3QR, UK}
\date{\today}

\begin{abstract}
Numerically optimised microwave pulses are used to increase excitation efficiency and modulation depth in electron spin resonance experiments performed on a spectrometer equipped with an arbitrary waveform generator. The optimisation procedure is sample-specific and reminiscent of the magnet shimming process used in the early days of nuclear magnetic resonance -- an objective function (for example, echo integral in a spin echo experiment) is defined and optimised numerically as a function of the pulse waveform vector using noise-resilient gradient-free methods. We found that the resulting shaped microwave pulses achieve higher excitation bandwidth and better echo modulation depth than the pulse shapes used as the initial guess. Although the method is theoretically less sophisticated than simulation based quantum optimal control techniques, it has the advantage of being free of the linear response approximation; rapid electron spin relaxation also means that the optimisation takes only a few seconds. This makes the procedure fast, convenient, and easy to use. An important application of this method is at the final stage of the implementation of theoretically designed pulse shapes: compensation of pulse distortions introduced by the instrument. The performance is illustrated using spin echo and out-of-phase electron spin echo envelope modulation experiments. Interface code between \emph{Bruker SpinJet} arbitrary waveform generator and \emph{Matlab} is included in versions 2.2 and later of the \emph{Spinach} library.
\end{abstract}


\maketitle

\section{Introduction}

A significant current problem in high-field electron spin resonance (ESR) spectroscopy is the difficulty of achieving uniform and quantitative signal excitation using microwave pulses \cite{SCHWEIGER01,BRUSTOLON09}. The greatest instrumentally feasible electron spin nutation frequency in wide-band ESR spectrometers at W-band (94 GHz) is about 50 MHz \cite{CRUICKSHANK09}. The corresponding \(\pi/2\) pulse is therefore 5~ns long, and the excitation bandwidth is around 200~MHz -- enough to affect a significant portion of many solid state ESR signals, but insufficient to excite such signals uniformly and quantitatively. The consequences of partial excitation include useful orientation selection effects \cite{POLYHACH07,BODE08,SCHIEMANN09}, but also reduced sensitivity and diminished modulation depth in two-electron dipolar spectroscopy \cite{THURNAUR80,SALIKOV92,TAN94}.

The time resolution of the best available microwave pulse shaping equipment is of the order of 20~ps\footnote{\emph{Keysight M8196A} 92 GSa/s Arbitrary Waveform Generator}. This work uses \emph{Bruker SpinJet} AWG with 0.625~ns time resolution -- it enables generation of shaped pulses with the (--3~dB) bandwidth of about 330~MHz and allows many broadband excitation schemes originally developed for nuclear magnetic resonance (NMR) spectroscopy \cite{WIMPERIS94} to be used with only minor modifications \cite{SPINDLER13,DOLL13,DOLL14,SCHOPS15,JESCHKE15}. Numerically designed ``optimal control'' microwave pulses \cite{KHANEJA05,FOUQUIERES11,GOODWIN16,GOODWIN17} are also possible \cite{SPINDLER12,KAUFMANN13}, but a complication specific to ESR is that the waveforms received by the sample are very different from those sent by the AWG -- the distortions introduced by the ESR instrument cannot be ignored \cite{SPINDLER12}.

\begin{figure}
\centering{\includegraphics{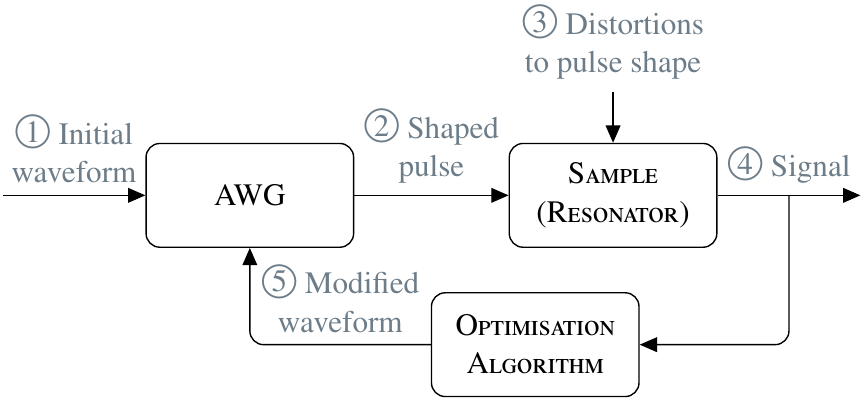}}
\caption{A schematic diagram of closed-loop feedback control: \circled{1} Initial waveform sent to the AWG. \circled{2} Waveform processed and sent to sample. \circled{3} Distortions from noise and hardware response. \circled{4} Sample excitation measured as a signal. \circled{5} New pulse shape calculated and sent to the AWG. \label{FIG-FeedbackControl}}
\end{figure}

One way around this is to introduce a transfer matrix or a response function that connects, under the linear response approximation, the ideal pulse emitted by the computer to the real pulse seen by the sample \cite{RIFE89,SPINDLER14}. The transfer matrix may be measured either by adding an antenna to the resonator \cite{SPINDLER12}, or by using a sample with a narrow ESR line to pick up the intensity of each frequency component \cite{KAUFMANN13}. Quasi-linear responses, such as phase variation across the excitation bandwidth, can be described with additional transfer matrices \cite{DOLL13}. A common procedure is to apply the experimentally measured response function at the pulse optimisation stage, to send the result out of the AWG, and to hope that a good rendering of the intended pulse arrives at the sample point. It usually does \cite{SPINDLER12,KAUFMANN13}, but the downside is that the linear response assumption is hard-wired into the process. Measuring the response function with a sufficient signal-to-noise ratio can be time-consuming. ESR resonators, particularly at high frequencies, also tend to have strongly sample-dependent response functions.

In this communication, we explore a different microwave pulse shape refinement strategy that does not use the linear response approximation. It relies instead on the possibility of repeating an ESR experiment hundreds of times per second, and recognises the fact, discussed in detail below, that microwave pulse shapes in ESR need very few discretisation points.

The method is known as ``feedback control'' \cite{FRANKLIN94,SKOGESTAD07,DOYLE13} and is illustrated schematically in FIG.~\ref{FIG-FeedbackControl}. It was originally proposed in the context of MRI \cite{LURIE85}, NMR \cite{LIU90}, and laser spectroscopy \cite{JUDSON92,BARDEEN97}. Its electron spin resonance adaptation is similar in its mathematical details to the well known (in the NMR circles) process of maximising the deuterium lock signal during the magnet shimming process \cite{DEROME13} -- a target variable is chosen and maximised, using a noise-resilient algorithm, with respect to the variables of interest. In relation to NMR pulse sequence optimisation, the method is known as ``direct spectral optimisation'' \cite{DEPAEPE03,ELENA04}.

In the ESR case explored in this work, the optimisation variables are either amplitudes of the microwave field at each time point, or parameters of the function describing the pulse shape. Improvements in excitation efficiency, spin-echo amplitude \cite{HAHN50} and signal modulation depth in out-of-phase electron spin-echo envelope modulation (OOP-ESEEM) \cite{TIMMEL98} experiments are demonstrated, at the instrument time cost not exceeding the time it used to take to auto-shim an NMR magnet -- minutes.

Although this method could be used to find an optimum pulse from a random guess, a more practical usage case would be to refine a pulse shape from some reasonably good starting point, such as a composite pulse or an open-loop optimal control solution \cite{GLASER15}, by compensating the inevitable distortions introduced by the ESR instrument \cite{HELLGREN13,EGGER14} in a way that does not rely on the linear response approximation.

\section{Feedback Control Optimisation}

\begin{figure}
\centering{\includegraphics{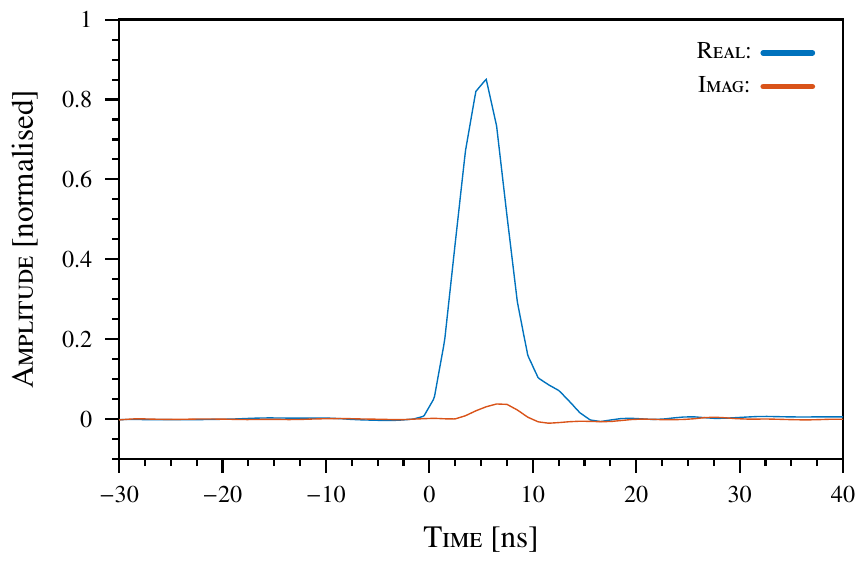}}
\caption{Pulse response function of our Bruker EleXSys II E580 spectrometer up to the TWT, measured by sending a large number of randomly generated waveforms (${\bf{x}}$) out of the AWG, recording the signals arriving at the transmitter monitor positioned just before the TWT (${\bf{y}}$), and solving the overdetermined system of ${\bf{y}} = {\bf{Px}}$ equations for the transfer matrix ${\bf{P}}$ using the SVD pseudoinverse procedure. The plot shows one of the columns of ${\bf{P}}$.\label{FIG-ResponseFcn}}
\end{figure}

\subsection{Arbitrary waveform generator interface}

The \emph{Bruker SpinJet} AWG used in this work has a time resolution of 0.625~ns, 14-bit amplitude resolution, 1.6~GS/s sampling rate, and \(\pm\)400~MHz range around the carrier frequency. The \emph{Bruker EleXSys II E580} ESR spectrometer has a 2~ns time base and, in combination with the AWG, resolves the time resolution mismatch by downsampling the pulse waveform onto a 1~ns increment time grid. The pulse response function up to the travelling wave tube (TWT) is shown in FIG.~\ref{FIG-ResponseFcn}.

\begin{figure}
\centering{\includegraphics{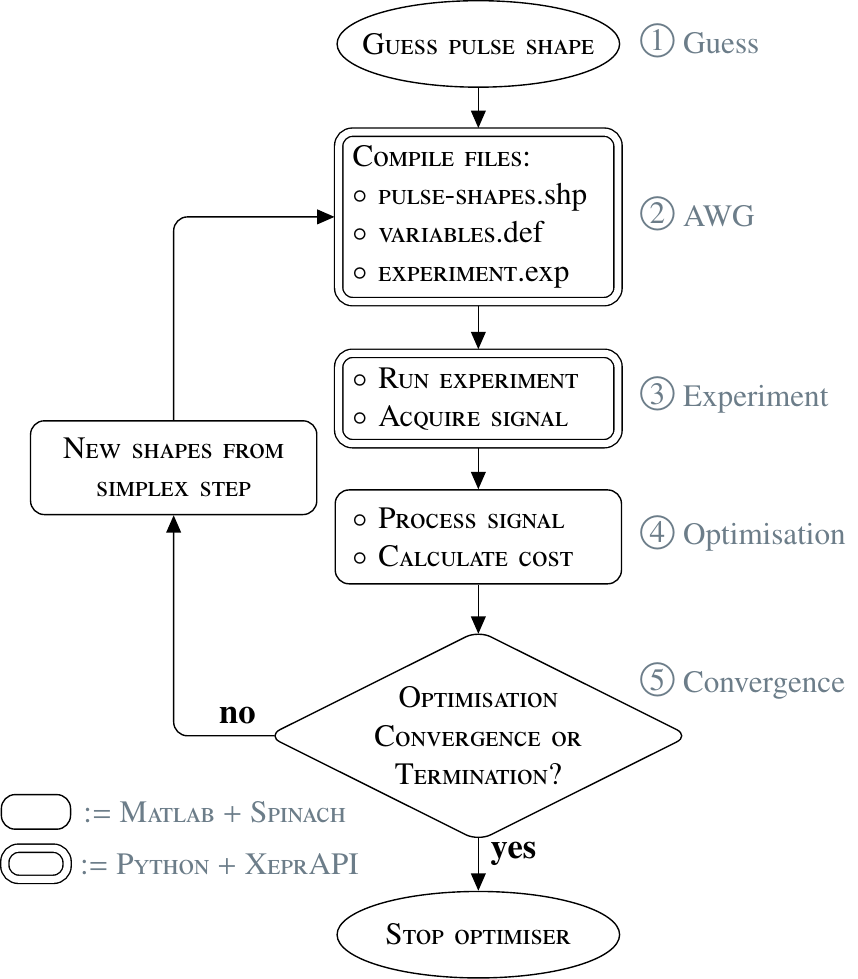}}
\caption{Software flow diagram: \circled{1} Initial guess provided by user. \circled{2} Waveform sent to AWG: load, show, and compile shape file, validate and compile PulseSPEL files. \circled{3} Run experiment, read signal data. \circled{4} Pass data to optimiser. \circled{5} If no optimisation convergence, calculate new pulse shape.}\label{FIG-SoftwareFlow}
\end{figure}

The software used in this work was written in-house, and has a flow of communication between \emph{Spinach} \cite{HOGBEN11} and \emph{Xepr Python} libraries, shown in FIG.~\ref{FIG-SoftwareFlow}. The master process runs in \emph{Matlab} and calls \emph{Xepr Python} functions as necessary to control the instrument. Experimental data is written by \emph{Xepr} into ASCII text files that are subsequently parsed by \emph{Matlab}. Optimisation restart capability is implemented using an MD5 hash table of the previously submitted experimental settings and outcomes \cite{GOODWIN16} -- an interrupted optimisation can therefore retrace its steps quickly without re-running previously executed experiments. 

To ensure that only feasible pulse shapes are sent from the software, the amplitude is folded into \([-1,+1]\) with the following mapping:
\begin{center}\includegraphics{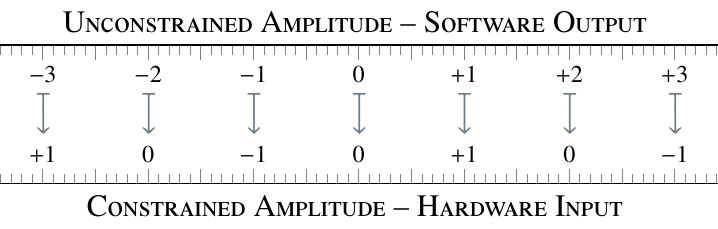}\end{center}
This is essentially a linear sawtooth map that takes \([-\infty,+\infty]\) into \([-1,+1]\) in a way that allows the optimisation algorithm to travel in \([-\infty,+\infty]\) and therefore removes the need to introduce constraints into the optimisation process.

\subsection{Optimisation method}

Numerical optimisation routines try to find extrema of the objective function supplied by the user \cite{GILL81,FLETCHER87,NOCEDAL06,GOODWIN17}. In ESR spectroscopy, the amplitude of a spin echo is a popular measure of sensitivity \cite{SCHWEIGER01,BRUSTOLON09} -- the stronger the echo, as measured by the integral of its real part, the better the excitation efficiency. Formally, the echo intensity may be defined as a norm of the real part $s(t)$ of the echo signal \cite{DOYLE13}, for example $L^1$-norm:
\begin{equation}
\big\| s(t) \big\|:=\int\limits_{0}^{T}\big| s(t)\big| \mathrm{d}t
 \label{EQ-Signal_1-Norm}
\end{equation}
where the integration is performed over the expected position of the echo (gained from an experiment with hard pulses). It is convenient to use a scaled objective function where the norm of the signal produced by the standard hard pulse experiment is used as a normalisation factor:
\begin{equation}
 \mathcal{Q}[s(t)]=\frac{\big\| s(t) \big\|}{\big\| r(t) \big\|}
 \label{EQ-Objective}
\end{equation}
Here, \(\mathcal{Q}\) is the functional to maximise, \(s(t)\) is the real part of the signal measured after running an experiment, and \(r(t)\) is the real part of the reference signal measured using hard pulses. \(\mathcal{Q}>1\) indicates a ``better'' echo than that from hard pulses alone, and \(0<\mathcal{Q}<1\) indicates a ``worse'' echo.

Numerical derivatives of noisy signals are unstable, and gradient-free optimisation strategies \cite{SWANN72,GILL81,FLETCHER87,NOCEDAL06,BRENT13,GOODWIN17} with a suitably scaled and bounded waveform are therefore to be preferred for feedback control optimisation. We found that a good choice is the Nelder-Mead algorithm \cite{NELDER65,LAGARIAS98} -- a member of the family of direct search methods known also as simplex methods \cite{SWANN72}, polytope methods \cite{GILL81} and \emph{ad hoc} methods \cite{FLETCHER87}. The method has the benefit of a smaller number of experimental evaluations per iteration compared to other gradient-free techniques, such as genetic algorithms \cite{JUDSON92,BRIF10} or simulated annealing \cite{METROPOLIS53}. However, convergence is not guaranteed, and is linear at best \cite{DENNIS87}. A desirable benefit of the Nelder-Mead algorithm is its tolerance to random noise on a smooth function \cite{GILMORE95,BORTZ98,KELLEY99}. In practical testing, we have found that a particular modification of the Nelder-Mead algorithm, called the multidirectional search method \cite{TORCZON89,DENNIS91,HIGHAM93,GOODWIN17}, works best. Its performance appears to be comparable to gradient descent algorithms; it is also designed to cope with multiple local minima \cite{GILMORE95}. A further useful property of the multidirectional search algorithm is that it has guaranteed convergence \cite{TORCZON91}.

\section{Materials and Methods}

ESR measurements were performed on a \emph{Bruker Biospin EleXSys II E580} spectrometer with a \emph{SpinJet} AWG based on an \emph{SPDevices SDR14} PCI board with a 0.625~ns time base. Samples were held at 85~K in an \emph{Oxford Instruments CF935O} cryostat under a flow of cold nitrogen gas, controlled by an \emph{Oxford Instruments Mercury} temperature controller. At X-band, the \emph{Bruker Biospin ER4118-MD5-W1} sapphire dielectric resonator with dimensions 5~mm ID, 10~mm OD and 13~mm height was used. The resonator was overcoupled for pulsed measurements to the quality factor of about 200. 

For spin echo optimisation, the test sample was 2.0~\(\mu\)M Finland trityl \cite{ardenkjaer1998epr} dissolved in a mixture of 30\% (by volume) glycerol-d\(_8\) and 70\% D\(_2\)O. 

\begin{figure}
\centering{\includegraphics[width=\columnwidth]{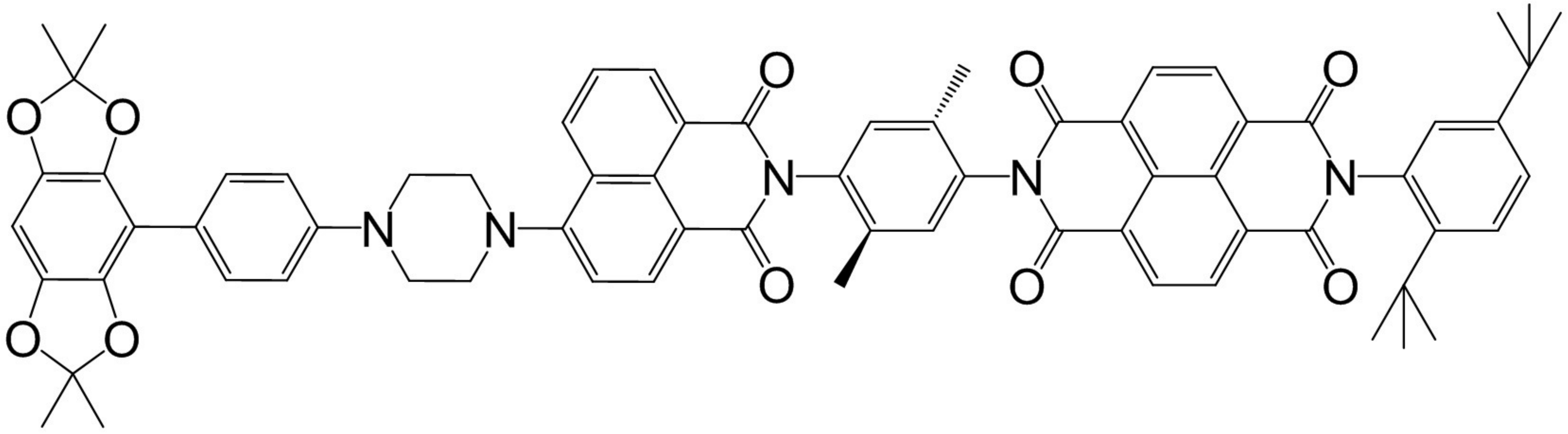}}
\caption{The sample used in OOP-ESEEM experiments on a photo-generated radical pair. The group that proposed this system \cite{CARMIELI09} used BDX tag for the biphenyl dioxolane derivative on the left, ANI tag for the 4-aminonaphthalene-1,8-imide (middle), and NI tag for the naphthalene-1,8-imide-4,5-imide (right), hence the overall BDXANINI tag for the construct. \label{FIG-BDXANINI}}
\end{figure}

OOP-ESEEM experiments used a sample of BDXANINI (FIG.~\ref{FIG-BDXANINI}), with the expected donor-acceptor distance of 26 \AA, similar to Sample 1 in \cite{CARMIELI09}, but with naphthalimide (NI) rather than phthalimide (PI) derivative as the acceptor, and 2,5-ditertbutylphenyl as the end group. BDXANINI was dissolved in 4-cyano-4\(^{\prime}\)-pentylbiphenyl (5CB) liquid crystal at a concentration of 0.2~mM and degassed in a 4~mm OD quartz tube using the freeze-pump-thaw method prior to flame sealing under vacuum. Prior to measurements, the sample was heated up to the isotropic phase and frozen. Photoexcitation of the BDXANINI sample was performed with a \emph{Continuum Surelite} Nd:YAG laser (7~ns pulses at 10 Hz at 1064~nm, 1~mJ per pulse, frequency tripled to 355~nm) attenuated with a \({\lambda}/{2}\) plate and depolarized. The beam was found to match the 5~mm cryostat window and was not manipulated further. Laser synchronisation was performed using a \emph{PatternJet II E580} board as an external trigger to a \emph{Stanford Research DG645} delay generator.

\section{Experimental Results}

\subsection{Feedback optimised spin echo}

\begin{figure}
\centering{\includegraphics{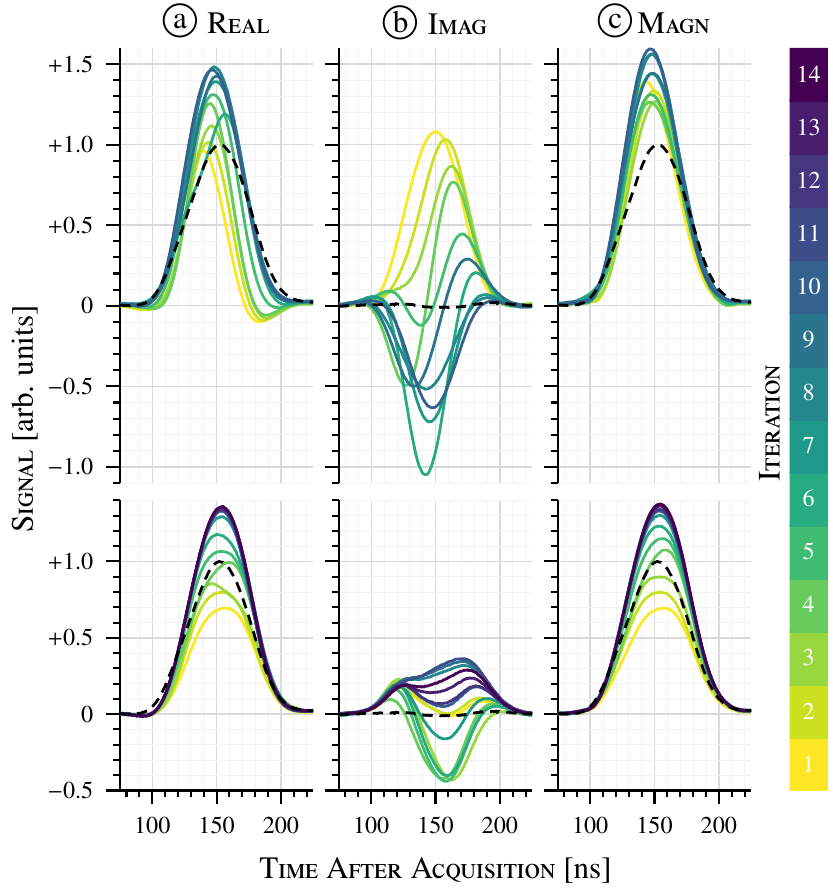}}
\caption{Optimisation process (colour-coded by iteration count) for a spin echo experiment performed on the trityl radical. The pulse waveform was optimised with respect to 11 discrete shape points (upper plots), and 21 discrete shape points (lower plots), with an overall $\pi$-pulse duration of 32~ns. The resulting real (a), imaginary (b), and magnitude (c) components of the echo signal are shown. The hard pulse reference signal is indicated by a dashed line.\label{FIG-EchoResults}}
\end{figure}

A two-pulse spin echo experiment \cite{HAHN50,ROWAN65}, consisting of an excitation pulse followed by an inversion pulse,
\begin{equation*}
  \overset{\makebox{16~ns}}{\underset{\text{hard pulse}}{\textsc{Excitation}}}  \xrightarrow[\text{delay}]{\makebox[1.5cm]{260~ns}}  \overset{\makebox{32~ns}}{\underset{\text{shaped pulse}}{\textsc{Inversion}}}  \xrightarrow[\text{delay}]{\makebox[1.5cm]{260~ns}}  \overset{\makebox{\ensuremath{s(t)}}}{\underset{\text{echo signal}}{\textsc{Acquisition}}}
\end{equation*}
was used to test the performance of the feedback control optimisation. The $L^1$-norm of the real part of the echo was maximised by varying the shape (both the in-phase and the quadrature component) of a 32~ns inversion pulse with an echo delay of 260~ns.

\begin{figure}
\centering{\includegraphics{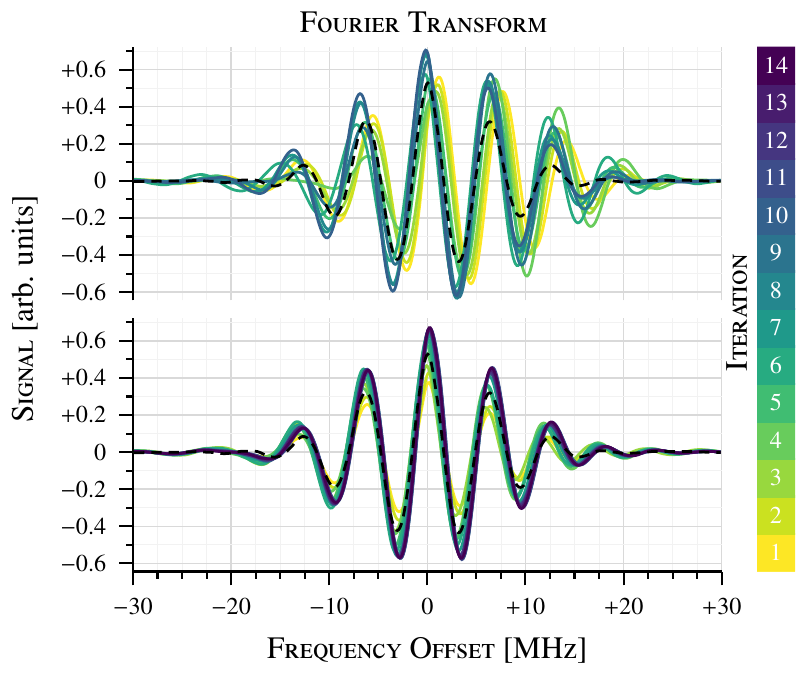}}
\caption{Optimisation process (colour-coded by iteration count) for a spin echo experiment performed on the trityl radical. The pulse waveform was optimised with respect to 11 discrete shape points (upper plots), and 21 discrete shape points (lower plots), with an overall $\pi$-pulse duration of 32~ns. Fourier transforms (after zero-filling and a sine bell apodization) of the corresponding echo signals from FIG.~\ref{FIG-EchoResults} are shown. The hard pulse reference signal is indicated by a dashed line. \label{FIG-FTEchoResults}}
\end{figure}

The initial condition was a random waveform. The echo signals at each iteration of the optimisation process are shown in FIG.~\ref{FIG-EchoResults} for the inversion pulse divided into 11 (10 \(\times\) 3.2~ns time slices) and 21 (20 \(\times\) 1.6~ns time slices) discrete points, interpolated linearly by the instrument. The Fourier transforms of these signals are shown in FIG.~\ref{FIG-FTEchoResults}. Both pulses converge to an echo better than that of the optimum hard pulse from a variety of random initial guesses. Predictably, the 21-point shape takes more iterations to converge.

It is clear from FIG.~\ref{FIG-EchoResults} that a significant improvement in the echo intensity is accompanied by the appearance of out-of-phase components. This is the consequence of the target functional in EQ.~\ref{EQ-Objective} placing no constraints on the imaginary part of the echo signal. This is intentional for a simple demonstration: a different choice of the figure of merit (for example, a difference between the norm of the in-phase part and the norm of the out-of-phase part) would suppress the out-of-phase component. A variety of other target functionals (for example, placing the magnetisation into a particular point on the Bloch sphere) and a discussion of their use may be found in the optimal control literature \cite{KHANEJA05,FOUQUIERES11,GOODWIN16,GOODWIN17}.

\begin{figure*}
\centering{\includegraphics{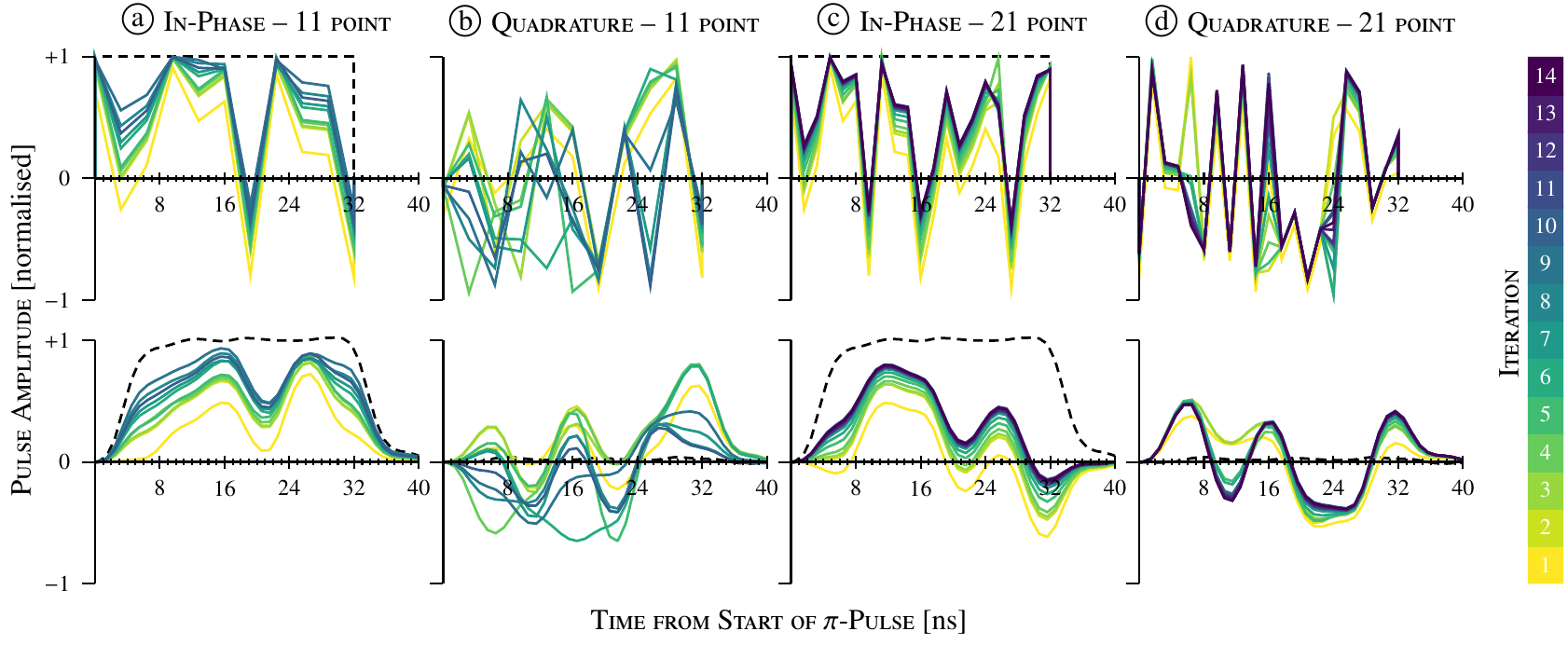}}
\caption{Optimisation process (colour-coded by iteration count) for the in-phase and quadrature parts of (a-b) 11-point and (c-d) 21-point pulse shapes producing the echoes in FIG.~\ref{FIG-EchoResults}. Lower plots show the convolution of the pulse shapes in the upper plots with the experimentally determined instrument response function. The hard pulse reference signal is indicated by a dashed line. \label{FIG-TransShapes}}
\end{figure*}

The pulse shapes at each iteration of the optimisation process are shown in FIG.~\ref{FIG-TransShapes}. Although it is possible to run the optimisation with more waveform discretisation points, this was not in practice found to produce any further improvement. This may be rationalised by inspecting the spectrometer response function in FIG.~\ref{FIG-ResponseFcn}, and a similar one published recently by the Prisner group \cite{SPINDLER12}. The width of the kernel is around 5~ns even before the TWT and the resonator, and therefore all finer details of the pulse waveform are lost in the convolution process. It is then to be expected that the nearby points in a 21-point 32~ns waveform would become correlated, and this would make the job more difficult for the optimisation algorithm. Given the width of the response function, a more reasonable discretisation is about one point every 3~ns -- almost exactly as in our 11-point shape shown in FIG.~\ref{FIG-EchoResults}, where the optimiser performs well and runs in good time.

The convolution of the 11-point and the 21-point optimum pulses from FIG.~\ref{FIG-EchoResults} with the experimentally measured response function for our \emph{EleXSys II E580} spectrometer is shown in FIG.~\ref{FIG-TransShapes}. An important conclusion is that the development of waveform shaping equipment in ESR spectroscopy should include improving the overall instrument response function -- there is little to gain from a sub-nanosecond AWG when the width of the response kernel, even without the amplifier or the resonator (FIG.~\ref{FIG-ResponseFcn}), is several nanoseconds.

\subsection{Feedback optimised OOP-ESEEM}

A reasonable measure of sensitivity in an OOP-ESEEM experiment \cite{TIMMEL98} is the modulation depth -- the difference between the initial amplitude of the echo, and its amplitude at the first minimum  \cite{DOLL15,DOLL16}. A norm of the difference between these echo signals is therefore one possible figure of merit:
\begin{equation}
\mathcal{Q}[s_1^{}(t),s_2^{}(t)]=\frac{\big\| s_1^{}(t) - s_2^{}(t) \big\|^{}}{\big\| r_1^{}(t) - r_2^{}(t) \big\|^{}}
 \label{EQ-EseemObjective}
\end{equation}
where \(s_{1}^{}(t)\) is the echo signal in an experiment with inter-pulse delay chosen to correspond to the first modulation maximum, \(s_{2}^{}(t)\) is the echo signal in an experiment with inter-pulse delay chosen to correspond to the first modulation minimum. Reference signals \(r_{1}^{}(t)\) and \(r_{2}^{}(t)\), produced by a sequence with hard pulses, are used for normalisation. 

In this case, we chose to optimise the excitation pulse of the photo-induced OOP-ESEEM sequence:
\begin{equation*}
{\underset{h\nu}{\textsc{laser}}} \rightarrow {\underset{\text{shaped pulse}}{\textsc{excitation}}} \xrightarrow[\text{delay}+n\Delta]{\makebox[1cm]{}} \overset{\makebox{}}{\underset{\text{hard pulse}}{\textsc{inversion}}}  \xrightarrow[\text{delay}+n\Delta]{\makebox[1cm]{}} \overset{\makebox{}}{\underset{\text{echo signal}}{\textsc{acquisition}}}
\end{equation*}

The photogenerated BDXANINI biradical has the first modulation maximum at 16~ns, and the first modulation minimum at 144~ns. These times were determined from an experiment with hard pulses, but it is also possible to allow these delays to vary during the optimisation.

The efficiency of the first pulse in ESEEM type sequences is a function of the bandwidth that the pulse is able to excite. Linear frequency sweep pulses \cite{DOLL16}, such as chirp \cite{BOHLEN90,BOHLEN93} and WURST (officially ``wideband, uniform rate, smooth truncation'' \cite{KUPCE95a,KUPCE95b}, but more likely because the polar plot in 3D resembles a sausage), are successful for broadband excitation in ESR \cite{SPINDLER12,SPINDLER13,DOLL13,DOLL14,DOLL16}, and it is therefore reasonable to optimise the parameters of those pulses instead of treating the entire waveform as a variable vector. 

The WURST pulse is defined by its amplitude and phase:
\begin{equation}
A(t)= 2\pi\sqrt{\frac{\omega_\text{bw}}{T}}\Bigg(1-\bigg|\sin^a{\bigg(\frac{\pi t}{T}\bigg)}\bigg|\Bigg),\qquad\varphi(t)= \pi\frac{\omega_\text{bw}}{T} t^2+\varphi_c^{}
\label{EQ-Wurst}
\end{equation}
where \(T\) is the pulse duration, \(\omega_\text{bw}\) is the excitation bandwidth, \(a\) is the sine power used for the amplitude envelope, and \(t\) runs over the interval \(\left[-T/2,+T/2\right]\). The amplitude in EQ.~\ref{EQ-Wurst} is normalised to produce a \(\pi\)-pulse. Other flip angles are obtained by changing the duration or the amplitude \cite{JESCHKE15,SPINDLER17}. 

\begin{figure}
\centering{\includegraphics{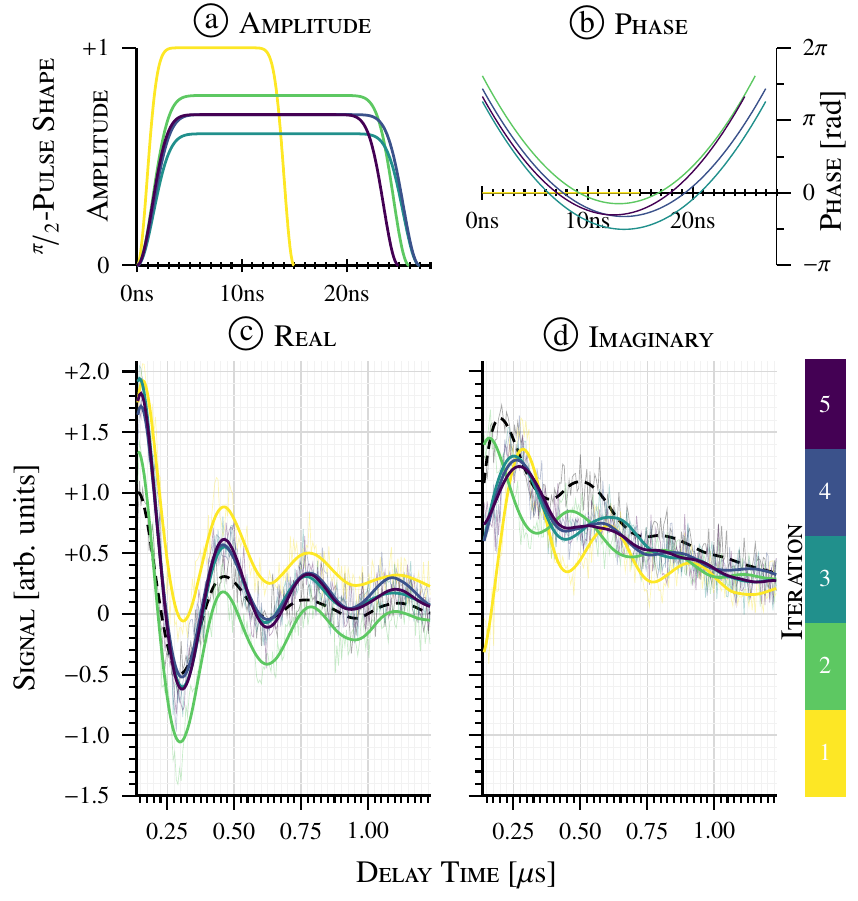}}
\caption{Modulation depth optimisation process (colour-coded by iteration count), showing (a-b) amplitude and phase of the excitation pulse, and (c-d) real and imaginary parts of the OOP-ESEEM signal. Thin lines show the raw data and bold lines are smoothed with a Savitzky-Golay filter \cite{SAVITZY64} (25-point cubic 1\(^\text{st}\) pass, 75-point quintic 2\(^\text{nd}\) pass). The hard pulse reference is indicated by a dashed line. The modulation depth is defined as the difference between the initial echo intensity and the intensity at the first minimum. \label{FIG-OopeseemMD}}
\end{figure}

Excitation pulse amplitude, duration, phase offset, and bandwidth were set as optimisation variables, and a phase cycle with \(\varphi_c^{}\in\{0,\pi/2,\pi,3\pi/2\}\) was set up. The initial guess was a 16~ns pulse with a zero phase offset, maximum amplitude, and a zero frequency sweep range, making it essentially a square pulse.

The small number of parameters makes the simplex method very efficient:  FIG.~\ref{FIG-OopeseemMD} shows an improvement in the modulation depth by over 50\% after just five iterations that ran in less than 3~minutes.

\subsection{Effect of instrument response function}

The excitation dynamics in the photo-generated BDXANINI biradical was simulated using \emph{Spinach} library \cite{HOGBEN11} for the corresponding two-electron spin system. Interaction parameters were those obtained by Wasielewski group \cite{COLVIN13} by fitting ESR data: the eigenvalues of \emph{g}-tensors are \(\big(2.0045, 2.0045, 2.0045\big)\) for the BDX\(^{+\bullet}\) donor and \(\big(2.0032, 2.0032, 2.0011\big)\) for the NI\(^{-\bullet}\) acceptor. The exchange coupling is 180~\(\mu\)T and the effective inter-electron distance is 26~\AA. 

The WURST pulse was applied numerically, and the fidelity calculated as the real part of the overlap between the final state of the system and the target state \(|\sigma\rangle\):
\begin{equation}
    \mathcal{F}=\text{Re}\langle\sigma|{\mathbf{P}}_N^{}{\mathbf{P}}_{N-1}^{}\cdots{\mathbf{P}}_2^{}{\mathbf{P}}_1^{}|\rho_0\rangle
    \label{EQ-Fidelity}
\end{equation}
where the initial state \(|\rho_0\rangle\) is the thermal equilibrium, and the propagator over an infinitesimally small time slice \(\Delta t\) is
\begin{equation}
    {{\bf{P}}_n} = \exp \left( { - i\left[ {{\omega _{\rm{o}}}{{\bf{H}}_{\rm{Z}}} + {A_n}\cos \left( {{\varphi _n}} \right){{\bf{H}}_{\rm{X}}} + {A_n}\sin \left( {{\varphi _n}} \right){{\bf{H}}_{\rm{Y}}}} \right]\Delta t} \right)
\end{equation}
where \(\omega_\text{o}\) is the resonance offset and \({\mathbf{H}}_{\rm{X,Y,Z}}\) are Cartesian spin operators. The destination state is in the XY plane with the phase matched to the offset in the following way:
\begin{equation}
    |\sigma(\phi)\rangle=\sin{(\phi)}|x\rangle - \cos{(\phi)}|y\rangle
\end{equation}
\begin{equation} 
    \phi(\omega_\text{o})=\frac{\pi T (\omega_\text{o}-\omega_\text{o}^2)}{\omega_\text{bw}}
\end{equation}

\begin{figure}
\centering{\includegraphics{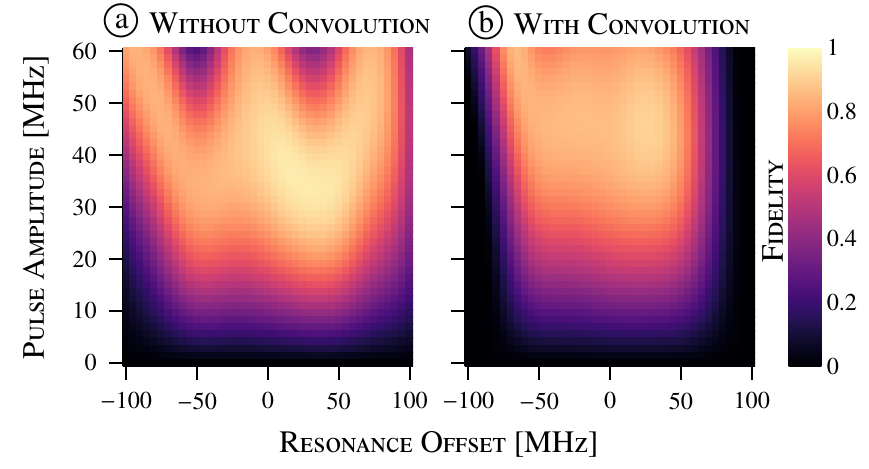}}
\caption{Robustness of the final pulse shape in FIG.~\ref{FIG-OopeseemMD} with respect to the instrument response convolution. Panel (a) shows the fidelity as a function of pulse amplitude and resonance offset for the theoretical WURST pulse shape with optimal parameters; panel (b) shows the fidelity achieved by a pulse that was convolved with the instrument response function shown in FIG.~\ref{FIG-TransShapes}. The fidelity is normalised to the maximum achievable value. \label{FIG-OopeseemRobust}}
\end{figure}

The considerable effect of the instrument response function on the performance of the WURST pulse is illustrated in FIG.~\ref{FIG-OopeseemRobust}, where the left hand side shows excitation efficiency in the absence of any distortions, and the right hand side includes the effect of the response function. A reduction in the tolerance to the resonance offset is apparent.

Because the response function is a hard to predict collective property of the console, the resonator, and the sample, some degree of feedback optimisation is likely to be required in any experiment that seeks to attain maximum possible performance. A combination of analytical design and open-loop control, with feedback control as the last stage, is currently viewed as the most promising strategy \cite{GLASER15}.

\section{Discussion}

An important question regarding the improvements seen in the applications above is about the source of those improvements: \emph{what does a numerically optimised pulse do that a shaped pulse did not?} This matter is discussed at length in the optimal control literature -- the following factors are pertinent:

\begin{enumerate}
\item{Because the optimum implies a zero gradient, optimised pulses are stable to first order with respect to variations in the optimised parameters, even in the presence of instrument distortions which may be significant (FIG.~\ref{FIG-TransShapes}).}
\item{Feedback control optimisation does not make the linear response assumption that is implicit in any method that relies on measuring the response function. An experimentally measured response function may also be noisy and thus a limiting factor in the optimisation.}
\item{A typical frozen glass ESR sample contains distributions in a variety of parameters: spin system orientation, interaction amplitudes (for example, due to conformational mobility), $B_1$ field strength, temperature, \emph{etc.} The pulse that comes out of a feedback optimisation does by definition have the best achievable performance, in the chosen class of functions, in the presence of all those distributions.}
\item{A numerically optimised pulse does not have to be adiabatic, and can explore the ``shortcuts to adiabaticity'' regime \cite{TORRONTEGUI13}. It is therefore possible to obtain a pulse reaching the target fidelity faster.}
\item{Feedback optimisation may be viewed as a generalisation of the pulse calibration process: more parameters than just duration and amplitude are now optimised, and pulse distortions by the instrument are hidden from view by the numerical optimisation algorithm.}
\end{enumerate}

\begin{figure}
\centering{\includegraphics{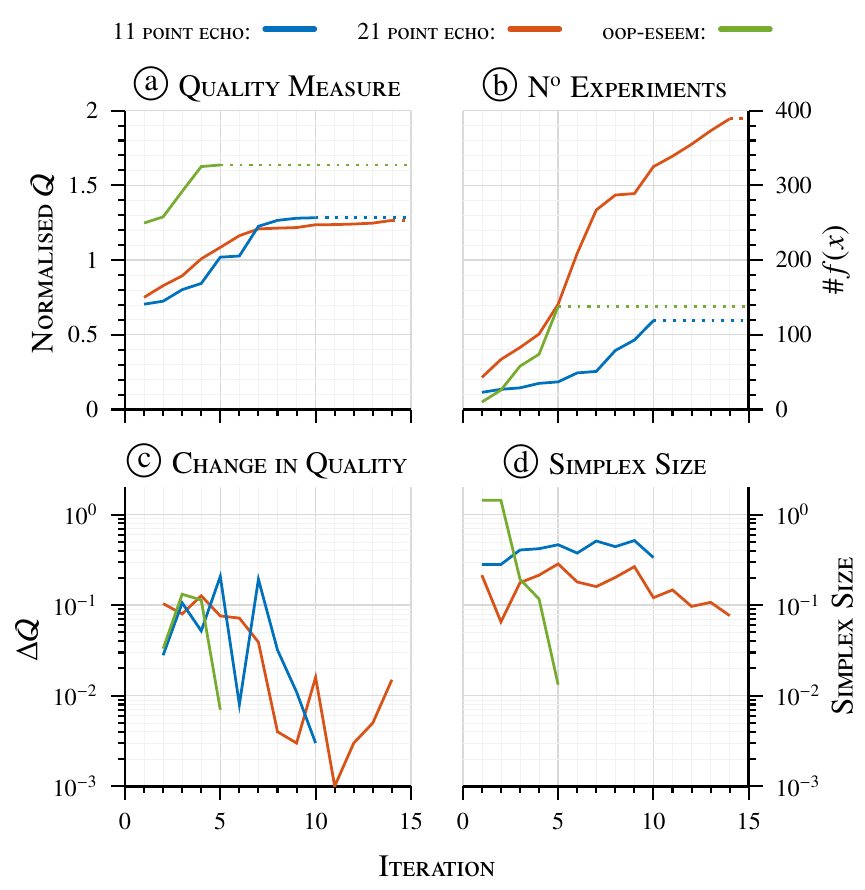}}
\caption{Convergence profiles for the pulse shape optimisations discussed in this work (echo pulse with 11 points, blue; echo pulse with 21 points, orange; OOP-ESEEM excitation pulse, green). The panels show (a) the quality metric as a function of the iteration count; (b) the cumulative number of ESR experiments performed by the instrument as a function of iteration count; (c) the change in the quality metric at each iteration; (d) the volume of the simplex at each iteration.}\label{FIG-ConvergenceData}
\end{figure}

The combination of these factors is responsible for the improved performance seen in FIGs.~\ref{FIG-EchoResults} and \ref{FIG-OopeseemMD}. Because so much of the ESR instrument response chain is sample-specific or hard to isolate, a feedback optimisation is nearly certain to result in significant performance improvements with respect to the target metric specified by the user. In common with other applications of optimal control theory, the optimisations started from different random initial guesses show similar convergence characteristics and final achievable fidelity, but can converge to very different final pulse shapes. FIG.~\ref{FIG-ConvergenceData} shows an example of the optimisation convergence behaviour for spin echo and OOP-ESEEM optimisations.

The limitations of the feedback control method involve the following factors:
\begin{enumerate}
\item{The initial signal-to-noise ratio. The S/N produced by the initial guess must be sufficiently high for the optimisation to be able to start. Simplex optimisers are resilient to the presence of noise, but practical experience indicates that a signal-to-noise ratio of at least 5 is required for the optimisation to make progress from the starting point.}
\item{The choice of the fidelity measure. This is a caveat common to all optimal control theory \cite{GLASER15}: the software will diligently maximise whatever it is given (for example, the real parts of the echoes in FIG.~\ref{FIG-EchoResults}), potentially at the expense of other important factors (like the imaginary parts of the same echoes) if they are not included into the fidelity measure. Taking care in the formulation of the optimisation target is the users' responsibility.}
\item{Interpretability. This is traditionally the weak spot of optimal control theory, with several recent papers specifically dedicated to the ways of finding out what the optimal pulse actually does and why \cite{KUPROV13,KOCHER14}; this is difficult.}
\item{Transferability. Feedback optimised pulse shapes are not, in general, transferable between different samples or different experiments. However, an optimal solution from a different experiment or a previous sample is usually an excellent initial guess.}
\end{enumerate}

Optimisation targets used in this work (echo intensity and ESEEM modulation depth) are not the only ones possible. Almost any instrument output parameter may be declared a target, and an optimal pulse shape or shapes obtained that maximises this target. Nor is the approach restricted to pulses: delays are also valid optimisation coordinates, as are any other systematically variable instrument settings.

\section{Conclusions and outlook}

Simple feedback control methods using gradient-free algorithms to optimise the performance of shaped pulses can lead to significant signal amplitude and modulation depth improvements in ESR experiments. The principal advantage of feedback control is that the linear response approximation, and the consequent need to measure the instrument response function, are avoided entirely.

The method presented here should be seen as complimentary to other pulse shape optimisation techniques already used in ESR \cite{SPINDLER12,SPINDLER13,KAUFMANN13,DOLL13,DOLL14,SCHOPS15,JESCHKE15,DOLL16,SPINDLER17}. Feedback control optimisation is beneficial as a fine-tuning step in the application of existing pulse sequences and optimal control methods \cite{HELLGREN13,EGGER14} because it provides a simple and general way to take into account hardware-specific and sample-specific variations in the pulse shape distortions introduced by the ESR instrument.

Our interface code between \emph{Matlab} and \emph{Bruker Xepr} is available in versions 2.2 and later of the \emph{Spinach} library \cite{HOGBEN11,GOODWIN17}.

\section*{Acknowledgements}
The authors are grateful to Arzhang Ardavan, Andrin Doll, Gunnar Jeschke, David Lurie, Burkhard Luy, and Thomas Prisner for insightful discussions, Michael Wasielewski for providing the sample material for OOP-ESEEM, and technical support of \emph{Bruker} through Patrick Carl and Peter H\"{o}fer. This work was supported by the EPSRC (EP/L011972/1, EP/J500045/1), and EU FP7 (297861/QUAINT).

\section*{References}

\bibliographystyle{elsarticle-num-names}
\bibliography{FBOptimCon_ESR_references}

\end{document}